\documentclass[aps,
               prx,
               showpacs,
               amssymb,
               superscriptaddress,
               twocolumn,
              nofootinbib,
              longbibliography]{revtex4-2}
\usepackage[utf8]{inputenc}
\usepackage{bbold}
\usepackage{bbm}
\usepackage{bm}
\usepackage{amsbsy}
\usepackage{amsthm}
\usepackage{amssymb}
\usepackage{amsfonts}
\usepackage{amsmath, mathtools}
\usepackage{dsfont}
\usepackage{graphicx}
\usepackage{epsfig}
\usepackage{epstopdf}
\usepackage{dsfont}
\usepackage{mathrsfs}
\usepackage{multibib}
\usepackage[dvipsnames]{xcolor}
\usepackage[colorlinks]{hyperref}
\makeatletter
\newcommand\org@hypertarget{}
\let\org@hypertarget\hypertarget
\renewcommand\hypertarget[2]{%
  \Hy@raisedlink{\org@hypertarget{#1}{}}#2%
  }
\makeatother
\usepackage[figure,table]{hypcap}
\usepackage{MnSymbol}
\usepackage{enumerate}
\usepackage{float}
\usepackage{comment}
\usepackage{braket}
\usepackage{tcolorbox}
\tcbuselibrary{theorems}
\newtcolorbox[blend into=tables]{smallboxtable}[3][]
{colback=mycolor!5, colframe=mycolor, float=ht, width=0.48\textwidth, lower separated=false, blend before title=colon hang,
title={#2}, label=#3 ,#1}
\newtcbtheorem{Definitions}{Definition}%
{colback=mycolor!5,colframe=mycolor,fonttitle=\bfseries,
left=4pt,
right=4pt,
top=5pt,
bottom=5pt}{defi}
\newtcbtheorem{Theorems}{Theorem}%
{colback=mycolor!5,colframe=mycolor,fonttitle=\bfseries,
left=4pt,
right=4pt,
top=5pt,
bottom=5pt}{theorem}
\definecolor{mycolor}{rgb}{0.122, 0.435, 0.698}
\usepackage{subfigure}
\usepackage{makecell}

\hypersetup{
	bookmarksnumbered,
	pdfstartview={FitH},
	citecolor={darkgreen},
	linkcolor={darkred},
	urlcolor={darkblue},
	pdfpagemode={UseOutlines}}
\definecolor{darkgreen}{RGB}{50,190,50}
\definecolor{darkblue}{RGB}{0,0,190}
\definecolor{darkred}{RGB}{238,0,0}
\usepackage{soul}

\newcommand{\ketbra}[2]{\ensuremath{|{#1}\rangle\!\langle{#2}|}}

\newcommand{\nl}{\ensuremath{\hspace*{-0.5pt}}}
\newcommand{\nr}{\ensuremath{\hspace*{0.5pt}}}

\newcommand{\subtiny}[3]{\ensuremath{_{\hspace{#1 pt}\protect\raisebox{#2 pt}{\tiny{$ #3$}}}}}
\newcommand{\suptiny}[3]{\ensuremath{^{\hspace{#1 pt}\protect\raisebox{#2 pt}{\tiny{$ #3$}}}}}

\newcommand{\tr}{\textnormal{Tr}}
\newcommand{\djj}{d\kern-0.4em\char"16\kern-0.1em}

\makeatletter
\renewcommand{\p@subsection}{}
\renewcommand{\p@subsubsection}{}
\makeatother

\newcommand{\Sys}{\ensuremath{_{\hspace{-0.5pt}\protect\raisebox{0pt}{\tiny{$\M{S}$}}}}}
\newcommand{\Mindex}[1]{\ensuremath{_{\hspace{-0.5pt}\protect\raisebox{0pt}{\tiny{$\M{M}_{#1}$}}}}} 
\newcommand{\SMi}{\ensuremath{_{\hspace{-0.5pt}\protect\raisebox{0pt}{\tiny{$\M{S}\M{M}_i$}}}}} 
\newcommand{\SM}{\ensuremath{_{\hspace{-0.5pt}\protect\raisebox{0pt}{\tiny{$\M{S}\hspace*{-1pt}\M{M}$}}}}} 
\newcommand{\Memo}{\ensuremath{_{\hspace{-0.5pt}\protect\raisebox{0pt}{\tiny{$\M{M}$}}}}} 

\newcommand{\M}[1]{\mathcal{#1}}

\newcommand{\id}{\mathds{I}}

\newcommand{\cmax}{C\subtiny{0}{0}{\mathrm{max}}}

\newcommand{\trho}{\tilde{\rho}}

\usepackage{orcidlink}

\begin{document}

\title{Unknown measurement statistics cannot be redundantly copied using finite resources}

\author{Tiago Debarba\,\orcidlink{0000-0001-6411-3723}}
\email{debarba@utfpr.edu.br}
\affiliation{Departamento Acad{\^ e}mico de Ci{\^ e}ncias da Natureza - Universidade Tecnol{\'o}gica Federal do Paran{\'a}, Campus Corn{\'e}lio Proc{\'o}pio - Paran{\'a} -  86300-000 - Brazil}
\affiliation{Atominstitut, Technische Universit{\"a}t Wien, Stadionallee 2, 1020 Vienna, Austria}
\author{Marcus Huber\,\orcidlink{0000-0003-1985-4623}}
\email{marcus.huber@tuwien.ac.at}
\affiliation{Atominstitut, Technische Universit{\"a}t Wien, Stadionallee 2, 1020 Vienna, Austria}
\affiliation{Institute for Quantum Optics and Quantum Information - IQOQI Vienna, Austrian Academy of Sciences, Boltzmanngasse 3, 1090 Vienna, Austria}
\author{Nicolai Friis\,\orcidlink{0000-0003-1950-8640}}
\email{nicolai.friis@tuwien.ac.at}
\affiliation{Atominstitut, Technische Universit{\"a}t Wien, Stadionallee 2, 1020 Vienna, Austria}

\begin{abstract}
Measurements can be viewed as interactions between a measured system and a pointer system that imprint information about the system on the pointer. For so-called unbiased interactions, the measurement statistics{\textemdash}the information corresponding to the diagonal of the system's initial density operator{\textemdash}with respect to a chosen measurement basis are transferred accurately, even if the pointer is initially in a mixed state. However, establishing measurement outcomes as objective facts also requires redundancy. We therefore consider the problem of unitarily distributing the outcome statistics to several pointers or quantum memories. We show that the accuracy of this process is limited by thermodynamic restrictions on preparing the memories in pure states: exact duplication of unknown outcome statistics is impossible using finite resources. For finite-temperature memories, we put forward a lower bound on the entropy production of the duplication process. This \emph{Holevo{\textendash}Landauer} bound demonstrates that the mixedness of the initial memory limits the ability to accurately transfer the same information to more than one memory component, thus fundamentally restricting the creation of redundancies while maintaining the integrity of the original information. Finally, we show how the outcome statistics can be recovered exactly in the classical limit{\textemdash}via coarse-graining or asymptotically as the number of subsystems of each memory component increases{\textemdash}thus elucidating how objective properties can emerge despite inherent imperfections.
\end{abstract}

\maketitle


\section{Introduction}\label{sec:introduction}

\noindent\emph{What does it take for information about a system to be established as an objective fact?} In quantum-mechanics terminology the superficial answer is that a measurement has to be performed. Despite varying views of what exactly constitutes a measurement, a minimal requirement is an interaction of the measured system with a measurement apparatus (or ``pointer"), resulting in a correlated joint state of system and pointer. Subsequently, the pointer itself can be probed by one or more observers, who read off the measurement result, be it from a digital display, an old-fashioned pointer needle, or some other indicator. We typically expect these results to be \emph{objective} in the sense that different observers agree on the specific outcome that they perceive the pointer to show. \emph{But how does such objectivity emerge from the correlated state of system and pointer?}\\[-2.5mm] 

Here, we take the point of view that the emergence of objective measurement outcomes requires two additional ingredients: first, the \emph{redundant} and \emph{robust}\footnote{Note the slight variation with respect to the recent use of the terms redundancy and consensus in~\cite{ChisholmInnocentiPalma2023}.} encoding of the correlations in the degrees of freedom of the pointer, and second, the interaction of the apparatus with an external environment, which leads to equilibration of the final pointer state: For a measurement outcome to be established as an objective fact it has to be copied to a sufficiently large fraction of a macroscopic apparatus. In turn, the latter equilibrates with its environment to yield an effectively irreversible transition~\cite{schwarzhans2023,EngineerRivlinWollmannMalikLock2024,FernandoMelo2024finite}. This ensures that the measurement outcome can be independently probed and thus agreed upon by different observers without disturbing the result.\\[-2mm]

\begin{figure}[t]
   \centering
   \includegraphics[scale=0.14]{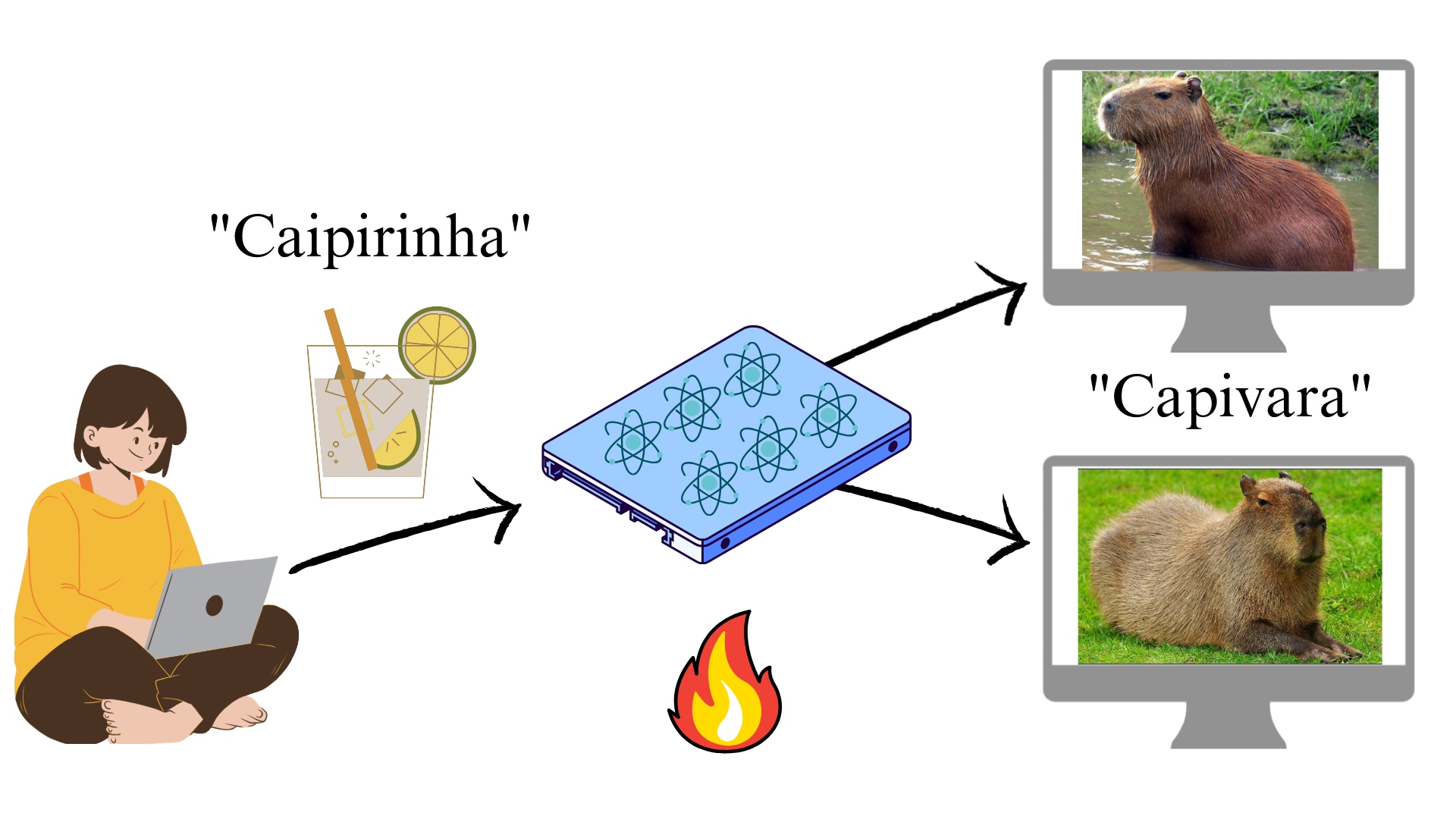}
   \vspace*{-3.5mm}
   \caption{Sonja
    ($\M{S}$, see~\cite{sonja}) wants to store a quantity of information $X$ (e.g., the word ``Caipirinha''), in a memory ($\M{M}$) composed of $N$ subsystems, such that multiple observers can access this information independently. However, if the memory has not been prepared in an ideal (pure) state, here meaning it is prepared in a thermal state, the ability to broadcast information is limited. Different memory subsystems thus contain imperfect information (e.g., the word ``Capivara'').}
   \label{fig: caipirinha}
\end{figure}

In this work we focus on the first aspect of the measurement process: creating redundancy. Specifically, we phrase the problem of redundant encoding as one of semiclassical broadcasting, i.e., the attempt to create multiple copies of information represented by (part of) the measured system's initial state in a \emph{memory} (the pointer or measurement apparatus). We refer to this process as semiclassical because we only wish to copy the outcome statistics represented by the diagonal elements (with respect to some chosen measurement basis) of the initial system density operator{\textemdash}which could be considered to be classical information, while the original broadcasting problem~\cite{BarnumCavesFuchsJozsaSchumacher1996} is concerned with copying the entire density operator. Yet, the process is not entirely classical, as we employ quantum-mechanical interactions. The process is also distinct from the original broadcasting problem in that the goal is to create two (or more) copies of the classical information but no conditions are made on the fate of the original quantum system. As we show here, the semiclassical broadcasting process is limited by the fact that we (fortunately and inescapably) live in a universe that is not at zero temperature and locally most macroscopic objects achieve (temporary) equilibrium best described by thermal states. However, establishing correlations from thermodynamic equilibrium has associated resource costs~\cite{HuberPerarnauHovhannisyanSkrzypczykKloecklBrunnerAcin2015,BruschiPerarnauLlobetFriisHovhannisyanHuber2015,VitaglianoKloecklHuberFriis2019,BakhshinezhadEtAl2019}, and perfect correlations are only achievable from zero temperature, which in turn requires diverging resources~\cite{TarantoBakhshinezhadEtAl2023}.
Thus, the first step of any measurement procedure is already plagued by inevitable imperfections~\cite{GuryanovaFriisHuber2018} that can only be mitigated by the suitable (costly~\cite{TarantoLipkaRodriguezBrionesBartosikPerarnauLlobetFriisHuberBakhshinezhad2024}) preparation of measurement devices.\\[-3mm] 

To arrive at this conclusion, we modify the measurement scenario usually modelled as an interaction between the measured system and a single memory to a more general case, where the information content of a quantum system is redundantly broadcast to multiple observers or memory components. Given that the use of finite resources introduces imperfections and limits the correlations between the quantum system and the memories, does it also impose restrictions on the ability to broadcast the informational content of the system? Our first main result answers this with a clear yes. The limitation lies in the number of memory components that can faithfully store the pertinent information about the system: just one. Therefore, a resource-theoretic approach of quantum physics not only imposes constraints on the ability to create perfect copies of the system itself, but also limits the capacity to replicate its informational content across multiple memories.\\[-3mm] 

The \emph{no-go theorem} we present here asserts that the mixedness of the initial memory fundamentally limits the ability to create redundancies while preserving the integrity of the copied information. This implies that entropy production in each memory component imposes constraints on the accessibility of the information broadcast to the memories. We cast this observation in terms of what we call the \emph{Holevo{\textendash}Landauer} bound, stating that, as entropy increases in the memory components, the broadcast information becomes less accessible, highlighting a trade-off between information redundancy and thermodynamic cost. However, we also identify two scenarios for mitigating finite-temperature effects in the classical limit. First, we show that post-processing several memory components allows one to recover the information exactly given sufficient knowledge of the initial memory and control over its interaction with the system. Second, ideal semiclassical broadcasting becomes possible asymptotically as the number of subsystems of each memory component diverges. Finally, we discuss implications for the emergence of objectivity.\\[-3mm] 

The remainder of this article is structured as follows: In Sec.~\ref{sec:non-ideal measurements} we give a brief overview of the framework for non-ideal measurements, before discussing the distinction between the original broadcasting problem and what we call semiclassical broadcasting in Sec.~\ref{sec:broadcasting vs semiclassical broadcasting}, and extending it to the non-ideal semiclassical broadcasting processes in Sec.~\ref{sec:non-Ideal broadcasting}. We then introduce the Holevo{\textendash}Landauer bound in Sec.~\ref{sec:holevo landauer}, and analyze different broadcasting scenarios that include jointly acting upon and post-processing information from multiple memories in Sec.~\ref{sec:variants of semiclassical broadcasting}. In Sec.~\ref{sec:classical limit and emergence of objectivity} we then turn to classical limits and the implications for the emergence of objectivity, and we conclude with a discussion in Sec.~\ref{sec:discussion}.


\section{Framework: Non-ideal measurements}\label{sec:non-ideal measurements}

{\noindent}A system $\M{S}$ contains information about a measurable quantity with possible values $x$ in the sense that outcomes of hypothetical ideal measurements on $\M{S}$ can be described by a random variable $X$ with probability distribution $\{p_x\}_{x}$. An actual measurement of the quantity can then be seen as an attempt to copy this information to a memory $\M{M}$ with suitable dimension~\cite{Dieks82,wooterzurek82}. However, this process of copying information is generally flawed: The information stored in the memory is represented by a different random variable $Y$ with probability distribution $\{q_{y}\}_{y}$. For classical systems, measurements can be thought of as not disturbing the system such that information can be freely copied by repeatedly interacting with the system.\\[-3mm]

The situation is different for quantum systems: Some measurements can reliably transfer the information represented by $\{p_x\}_{x}$ to the memory, so that $\{q_y\}_{y} =\{ p_x\}_{x}$, but may disturb the original system, leaving it to now contain information $\{\tilde{p}_x\}_{x}\neq \{p_x\}_{x}$. Following~\cite{GuryanovaFriisHuber2018}, these measurements are called \emph{unbiased}, while \emph{non-invasive} measurements do not disturb but may not correctly transfer the original information, $\{\tilde{p}_x\}_{x}= \{p_x\}_{x}\neq\{q_y\}_{y}$, as we will explain in more detail shortly. Whether a measurement falls into one of these categories depends not only on the specific interaction but also on the initial state of the measurement apparatus (here, $\M{M}$). In particular, \emph{ideal measurements}, defined as being both unbiased and non-invasive, can only be realized by unitary interactions if the initial state of $\M{M}$ is pure. Yet, the third law of thermodynamics prevents the exact preparation of pure states with finite resources (time, energy, and complexity of the involved apparatus and of the control over it, see~\cite{TarantoBakhshinezhadEtAl2023}) implying that all practical measurements are non-ideal, and can thus be \emph{unbiased} or \emph{non-invasive}, but not both simultaneously.\\[-2.5mm]  


Let us now make these statements more precise: We assume that the system $\M{S}$ and the memory $\M{M}$ are initially uncorrelated and hence in a product state of an in principle unknown state $\rho\Sys$ and a known state $\sigma\Memo$, respectively. A unitary interaction $U$, resulting in the joint final state
\begin{align}
    \tilde{\rho}\SM = U (\rho\Sys\otimes\sigma\Memo) U^{\dagger}\,,
\end{align}
can copy information encoded in the system to the memory such that, for a chosen basis $\{\ket{x}\}_{x}$ of~$\M{S}$ and set of orthogonal projectors $\{\Pi_x\}_x$ on~$\M{M}$, one ideally has \emph{unbiasedness}, 
\begin{align}
    \tr(\rho\Sys\,\ketbra{x}{x})  &=\,\tr(\tilde{\rho}\SM\,\id\Sys\otimes\Pi_x)\ \forall x\ \text{and}\ \forall \rho\Sys\,, 
    \label{eq:unbiasedness}
\end{align}
while not disturbing the system (\emph{non-invasiveness}) in the sense that 
\begin{align}
    \tr(\rho\Sys\,\ketbra{x}{x})=\tr(\tilde{\rho}\SM\,\ketbra{x}{x}\otimes\id\Memo)\ \forall x\ \text{and}\ \forall \rho\Sys\,.
    \label{eq:non-invasiveness}
\end{align}
The correlation between the post-interaction states of the system and memory can be captured by the quantity 
\begin{align}
    C_{U}(\rho\Sys,\sigma\Memo) &=\,\sum_x \tr(\tilde{\rho}\SM\,\ketbra{x}{x}\otimes\Pi_x)\,, 
\end{align} 
and we say the information is copied \emph{faithfully} if there exists a unitary $U$ such that $C_{U}=1$. 
In combination with either unbiasedness or non-invasiveness, \emph{faithfullness} implies the respective third property~\cite{GuryanovaFriisHuber2018}. 
However, for mixed initial states of $\M{M}$ with $\operatorname{rank}(\sigma\Memo)>r$,  
where $r=d\Memo/d\Sys$ is the ratio of the dimensions $d\Memo \geq d\Sys$ and $d\Sys$ of the memory $\M{M}$ and system $\M{S}$, respectively, 
the correlations represented by $C_{U}$ are limited to 
\begin{align}\label{eq:cmax}
    \cmax &=\, \max\limits_{U} \,C_{U}(\rho\Sys,\sigma\Memo)\,<\,1\,.
\end{align}
When $\cmax<1$, interactions can be either unbiased or non-invasive, but not both. There are also interactions that are neither unbiased nor non-invasive, but below, we focus on unitaries that have one of these properties and are also \emph{maximally correlating}, i.e., achieve $C_U=\cmax$.\\[-2mm]

To illustrate these concepts, let us consider the pertinent example of a memory prepared in a Gibbs state, $\sigma\Memo=\tau\Memo$, with
\begin{align}
    \tau\Memo(\beta) &=\,
    \exp(-\beta H\Memo)/\mathcal{Z}\Memo,
\end{align}
where $H\Memo$ is the memory Hamiltonian, $\beta=(k\subtiny{0}{0}{\mathrm{B}}T)^{-1}$ is the inverse temperature, $k\subtiny{0}{0}{\mathrm{B}}$ is the Boltzmann constant, and $\mathcal{Z}\Memo = \tr(\exp(-\beta H\Memo))$ is the partition function. 
The eigenvalues of $H\Memo$ can be grouped into $d\Sys$ disjoint sets $\{E\suptiny{0}{0}{(l)}\subtiny{0}{0}{k}\}_{k=0,1,\ldots,r}$ labelled by $l=0,\ldots,d\Sys-1$, 
such that $H\Memo=\sum_{k=0}^{r-1}\sum_{l=0}^{d\Sys -1}E\suptiny{0}{0}{(l)}\subtiny{0}{0}{k}\ketbra{E\suptiny{0}{0}{(l)}\subtiny{0}{0}{k}}{E\suptiny{0}{0}{(l)}\subtiny{0}{0}{k}}$. 
With this notation, the initial memory state can, for later convenience, be written as a sum of contributions from different subspaces, $\tau\Memo = \sum_y A_{0,y}$, with 
\begin{align}
    A_{0,y} &=\,\frac{1}{\M{Z}\subtiny{0}{0}{M}}\sum_{l=0}^{r-1} \exp\left(-\beta E\suptiny{0}{0}{(l)}\subtiny{0}{0}{y}\right)\ketbra{E\suptiny{0}{0}{(l)}\subtiny{0}{0}{y}}{E\suptiny{0}{0}{(l)}\subtiny{0}{0}{y}}\,.
    \label{eq:app_Global_M_matrices}
\end{align}
Moreover, for $\Pi_x = \sum_{l=0}^{r-1}\ketbra{E\suptiny{0}{0}{(l)}\subtiny{0}{0}{x}}{E\suptiny{0}{0}{(l)}\subtiny{0}{0}{x}}$, the quantity $\cmax$ is given by the sum of the $r$ largest eigenvalues of the initial memory state, 
\begin{align}
    \cmax  &=\,\frac{1}{\M{Z}\subtiny{0}{0}{M}}\sum_{l=0}^{r-1} \exp\left(-\beta E\suptiny{0}{0}{(l)}\subtiny{0}{0}{0}\right)\,.
    \label{eq:app_max_faithfulness}
\end{align}

\begin{figure}[t]
\begin{center}
\includegraphics[width=1.22\linewidth,trim={2.5cm 0.0cm 0cm 0.0cm},clip]{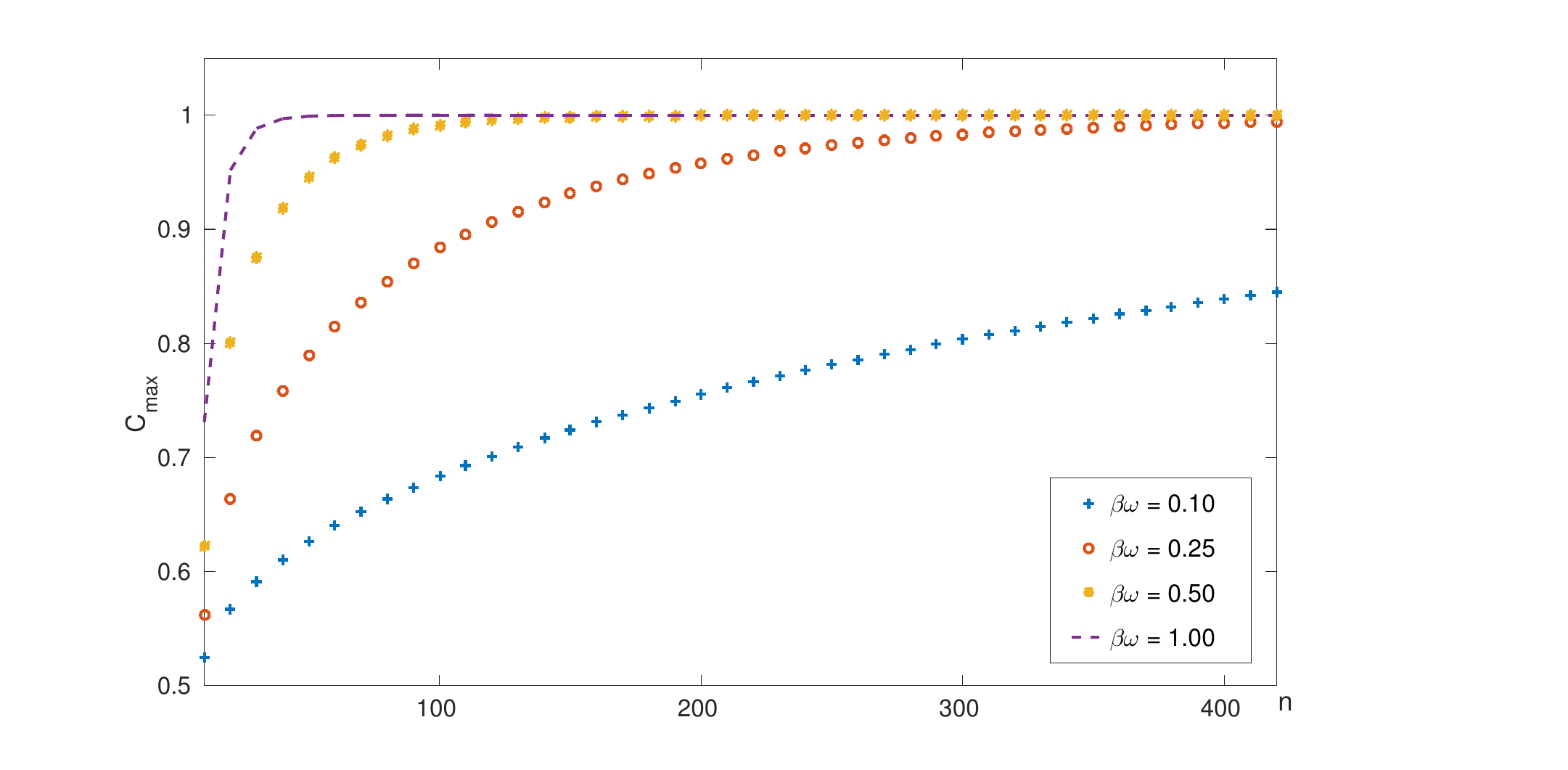}
\vspace*{-10mm}
\end{center}
    \caption{Plot illustrating the convergence of $\cmax$, for $\beta\omega =0.1, 0.25, 0.50, 1.00$, as a function of the number of qubits $n = [1,410]$ in each memory component. Each of the qubits is assumed to be described by the Hamiltonian $H\Memo\suptiny{0}{0}{(i)} = \omega\,\ketbra{1}{1}$, and initialized in a thermal state $\tau\Memo\suptiny{0}{0}{(i)} = (\ketbra{0}{0} + \exp(-\beta\omega)\ketbra{1}{1})/\M{Z}\Memo\suptiny{0}{0}{(i)}$, $\forall i=\{1,n\}$. In this example the \emph{maximal faithfulness} is given by  $\cmax = \sum_{m=0}^{(n-1)/2}\binom{n}{m}e^{-\beta m \omega}/\M{Z}\Memo^{n}$.}
    \label{fig:cmax}
\end{figure}

An example of a unitary acting jointly on the system and an initially thermal $n$-qubit pointer, and that is both unbiased (and hence invasive) and maximally correlating (achieving $C_U=\cmax$) is discussed in detail in~\cite[Sec.~A.VII]{GuryanovaFriisHuber2018}. Here, let us instead discuss an example of a unitary (again acting jointly on the system and an initially thermal $n$-qubit pointer) that is non-invasive and maximally correlating. In Fig.~\ref{fig:cmax}, we illustrate the convergence of $\cmax$ to~$1$ as the number $n$ of qubits in the memory increases while fixing the initial inverse temperature~$\beta$ and Hamiltonian $H=\omega\,\ketbra{1}{1}$ (with $\hbar=1$) for each qubit.\\[-2.5mm]  

A specific realization of a non-invasive unitary that achieves~$\cmax$ for the set $\{\ketbra{x}{x}\otimes\Pi_y\}_{x,y=0}^{d\Sys-1}$ is $U = \sum_x \ketbra{x}{x}\otimes V_x$, where for $x,y=0,1,\ldots, d\Sys-1$ the unitary operations $V_x$ act on $A_{0,y}$ as 
\begin{equation}\label{eq: v_x}
    A_{x,x\oplus y} = V_x A_{0,y} V_x^{\dagger}\,,
\end{equation}
where $x\oplus y = (x+y)\operatorname{mod}(d\Sys)$, and the resulting matrices have the explicit form
\begin{subequations}\label{eq:M matrices definition}
\begin{align}
A_{x,x} &=\frac{1}{\M{Z}\subtiny{0}{0}{M}}\sum_{l=0}^{r-1} \exp\left(-\beta E\suptiny{0}{0}{(l)}\subtiny{0}{0}{0}\right)\ketbra{E\suptiny{0}{0}{(l)}\subtiny{0}{0}{x}}{E\suptiny{0}{0}{(l)}\subtiny{0}{0}{x}}\,,\\
A_{x,x\oplus y} &=\frac{1}{\M{Z}\subtiny{0}{0}{M}}\sum_{l=0}^{r-1} \exp\left(-\beta E\suptiny{0}{0}{(l)}\subtiny{0}{0}{y}\right)\ketbra{E\suptiny{0}{0}{(l)}\subtiny{0}{0}{x\oplus y}}{E\suptiny{0}{0}{(l)}\subtiny{0}{0}{x\oplus y}}\,.
\end{align}
\end{subequations}
The information copied to the memory will depend on the reordering of $A_{x,x\oplus y}$ over the subspaces spanned by $\ketbra{x}{x}\otimes\Pi_y$~\cite{GuryanovaFriisHuber2018}.
The traces of the matrices in Eq.~\eqref{eq:M matrices definition} satisfy 
\begin{subequations}\label{eq: max correlation}
\begin{align}
a_{x,x} &=\, \tr(A_{x,x}) \,=\, \cmax\,,  \label{eq: max correlation - correlated}\\
\sum_{y=1}^{d\Sys-1} a_{x,x\oplus y} &=\, 
\sum_{y=1}^{d\Sys-1}\tr(A_{x,x\oplus y})\,=\,
1-\cmax\,\label{eq: max correlation - anti}
\end{align}
\end{subequations}
for all $x\in[0,d\Sys-1]$. 
As the set of the permutation unitaries $V_x$ is not unique, we may, for example, further impose that the transformation on the memory components is unital. 
In this case the state of the memory remains unchanged if $T\rightarrow \infty$ or 
$H\Memo=0$ and one may check that this implies 
$\sum_{x}a_{x,y}  = \sum_y a_{x,y} = 1$~\cite{DebarbaManzanoGuryanovaHuberFriis2019}.


\section{Broadcasting Versus Semiclassical Broadcasting}\label{sec:broadcasting vs semiclassical broadcasting}

{\noindent}Before we proceed with the analysis of non-ideal semiclassical broadcasting, it is crucial to clearly lay out the technical distinction between the well-known broadcasting problem~\cite{BarnumCavesFuchsJozsaSchumacher1996} and what we here call (redundant) semi-classical broadcasting. The term broadcasting has previously been used in the context of quantum information to signify the extension of the no-cloning theorem~\cite{Dieks82,wooterzurek82} to mixed states~\cite{BarnumCavesFuchsJozsaSchumacher1996}, sometimes phrased in terms of optimal cloning machines~\cite{GisinMassar1997,Werner1998}, for a review see~\cite{ScaraniIblisdirGisinAcin2005}. The central result of this programme is that quantum channels that can duplicate unknown (pure or mixed) states exactly do not exist. In particular, there is no unitary $U$ such that 
\begin{align}
    U\,(\rho\Sys\otimes\sigma\Memo)\,U^{\dagger}\,=\,\rho\Sys\otimes\rho\Memo
    \label{eq:original broadcasting}
\end{align} 
for a fixed state $\sigma\Memo$ independently of the input state $\rho\Sys$. Here it is interesting to note that, viewed as a measurement procedure in the sense that is discussed in the previous section, a successful broadcasting process would by definition be unbiased and non-invasive according to Eqs.~(\ref{eq:unbiasedness}) and~(\ref{eq:non-invasiveness}), respectively. Following~\cite{GuryanovaFriisHuber2018}, this would  imply a faithful measurement, $C_U=1$, but for arbitrary states $\rho\Sys$, the (uncorrelated) product state $\tilde{\rho}\SM=\rho\Sys\otimes\rho\Memo$ will not satisfy $C_U=\sum_x \tr(\tilde{\rho}\SM\,\ketbra{x}{x}\otimes\Pi_x)=1$, thus providing an alternative way of proving that the original, ideal broadcasting process is generically impossible.\\[-2mm]

Nevertheless it is possible to clone/broadcast the classical information content of the quantum state~\cite{Barnum2007}, represented by the diagonal of the density operator with respect to a fixed measurement basis, if one relaxes the condition of Eq.~(\ref{eq:original broadcasting}) to allow more general, correlated final states $\tilde{\rho}\SM$ with the property that the marginals $\tilde{\rho}\Sys=\tr\Memo(\tilde{\rho}\SM)$ and $\tilde{\rho}\Memo=\tr\Sys(\tilde{\rho}\SM)$ match the original system state, i.e., $\tilde{\rho}\Sys=\tilde{\rho}\Memo=\rho\Sys$. Specifically, for a given classical state $\rho\Sys = \sum_x p_x \,\ketbra{x}{x}$, there exists a unitary of the form $U = \sum_{x}\ketbra{x}{x}\otimes U_{x}$, with $U_{x}\ket{y} = \ket{x\oplus y}$, such that by setting $\sigma\Memo=\ketbra{0}{0}$ we have
\begin{align}
    \tilde{\rho}\SM \,=\,U\,(\rho\Sys\otimes\ketbra{0}{0}\Memo)\,U^{\dagger}
    &=\,\sum_x p_x\, \ketbra{x}{x}\Sys\otimes\ketbra{x}{x}\Memo\,,
    \label{eq:faithfull relaxed broadcast}
\end{align} 
a classically perfectly correlated state whose reduced states both match~$\rho\Sys$.\\[-2mm]

But also in this relaxed broadcasting scenario, successful broadcasting implies both unbiasedness and non-invasiveness, which together imply faithfulness,
\begin{align}
\sum_x \tr(\ketbra{x}{x} \otimes \Pi_x\,\trho\SM) \,=\, 1\,,
\label{eq: classical state broadcasting}
\end{align}
for $\Pi_x = \ketbra{x}{x}$. And, indeed, for an initially pure state of the memory, this is exactly what one achieves, as we see, for instance, in Eq.~(\ref{eq:faithfull relaxed broadcast}). But also for a mixed initial state of the memory the faithfullness condition in Eq.~(\ref{eq: classical state broadcasting}) implies that $\trho\SM$ has support only in a $d\Sys$-dimensional subspace of $\M{H}\Sys\otimes\M{H}\Memo$, and is hence rank deficient. Yet, if the memory is prepared using finite thermodynamic resources, $\rho\Memo$ must have full rank, and since the relations above must hold for any $\rho\Sys$, including states of full rank, also $\trho\SM$ must have full rank, in contradition to Eq.~(\ref{eq: classical state broadcasting}). The rank of the initial memory state thus restricts the ability to broadcast even classical information via unitary interactions, even when relaxing the original broadcasting scenario to only demanding matching marginals, resulting in a value lower than one for the left-hand side of Eq.~\eqref{eq: classical state broadcasting}. In short, finite resources prevent the memory from being rank-deficient, which is necessary for perfect broadcasting of classical states.\\[-2mm]

Here, we are interested in a scenario that we call \emph{semiclassical broadcasting} that is subtly but crucially different from the (relaxed) original broadcasting problem and hence is not covered by the rank argument above: We ask if it is possible to create multiple copies of the classical information corresponding to the diagonal elements of the initial system state in multiple memories, but we do not impose any conditions on the final state of the system. We cast this in terms of the following technical definition:\\

\begin{Definitions}{\rm Ideal semiclassical broadcasting}{ideal broadcasting}
    Let $\M{S}$ be a quantum system and $\M{M}$ be a quantum memory with $N$ components $\M{M}_{i}$ such that $\mathcal{H}\subtiny{0}{0}{\M{M}}=\bigotimes_{i=1}^{N}\mathcal{H}\subtiny{0}{0}{\M{M}_{i}}$, with initial states $\rho\Sys$ and $\rho\subtiny{0}{0}{\M{M}}$, respectively.
    A procedure that maps $\rho\Sys\otimes\rho\subtiny{0}{0}{\M{M}}$ to $\tilde{\rho}\subtiny{0}{0}{\M{SM}}$ is said to realize \textbf{ideal semiclassical broadcasting} from $\M{S}$ to the components $\M{M}_{i}$ with respect to the quantity $\hat{X}=\sum_x x\, \ketbra{x}{x}$ if there exist sets of orthogonal projectors $\Pi_x\suptiny{0}{0}{(i)}$ on $\mathcal{H}\subtiny{0}{0}{\M{M}_{i}}$ with $\Pi_{x}\suptiny{0}{0}{(i)}
    \Pi_{x'}\suptiny{0}{0}{(i)}
    =\delta_{xx'}\Pi_{x}\suptiny{0}{0}{(i)}$ such that
\begin{align}
    p_{x}  \,=\,  \bra{x}\rho\Sys\ket{x}\,=\,\tr(\tilde{\rho}\subtiny{0}{0}{\M{S\nl M}}\,\Pi_{x}\suptiny{0}{0}{(i)})\,=\,q_{x}\suptiny{0}{0}{(i)},
    \label{eq:ideal broadcasting}
\end{align}
    for all $x$ and $i$, and for all $\rho\Sys$.
\end{Definitions}


\section{Non-ideal semiclassical broadcasting.}\label{sec:non-Ideal broadcasting}

In the language used in Sec.~\ref{sec:non-ideal measurements}, ideal semiclassical broadcasting corresponds to demanding unbiasedness for each memory component. In particular, for single-component memories, Definition~\ref{defi:ideal broadcasting} coincides with that of an unbiased interaction~\cite{GuryanovaFriisHuber2018}. At the same time, Definition~\ref{defi:ideal broadcasting} does not require non-invasiveness, and hence does not imply the equivalent of faithfullness, Eq.~(\ref{eq: classical state broadcasting}). The previous rank argument hence does not rule out ideal semiclassical broadcasting. And, in particular, for $N=1$ ideal broadcasting is possible as long as the dimension of the memory is at least as large as that of the system, i.e., if $d\subtiny{0}{0}{\M{M}}:=\dim(\mathcal{H}\subtiny{0}{0}{\M{M}})\geq\dim(\mathcal{H}\Sys)=:d\Sys$. In a more general scenario with multiple memory components, it is clear that ideal semiclassical broadcasting requires the latter condition to hold separately for each component, $d\subtiny{0}{0}{\M{M}_{i}}\geq d\Sys$ for all $i=1,\ldots,N$, so that each memory component can in principle store the information in question. However, as we will see below, it also becomes evident that this is only a necessary and not a sufficient condition for ideal semiclassical broadcasting, which we will phrase in the following theorem.\\[-1.5mm] 

\begin{Theorems}{No ideal semicl. broadcasting}{no-go}
Ideal semiclassical broadcasting to multi-component memories ($N\geq2$) is impossible with finite resources: Specifically, no procedure represented by a joint unitary on the system $\M{S}$ and any finite-dimensional ($d\subtiny{0}{0}{\M{M}}<\infty$) memory $\M{M}$ that is initially described by a density operator $\rho\subtiny{0}{0}{\M{M}}$ with full-rank (in particular, a thermal state of a Hamiltonian with finite energy gaps) can realize ideal semiclassical broadcasting.
\end{Theorems}

\begin{proof}[Proof.]
We assume that ideal semiclassical broadcasting is possible for a two-component memory, $N=2$, and show that this leads to a contradiction. The two memory components can have different Hilbert spaces and different initial states $\rho\subtiny{0}{0}{\M{M}_{1}}$ and $\rho\subtiny{0}{0}{\M{M}_{2}}$, but both have full rank, and their von Neumann entropies are hence some non-zero constants, without loss of generality we assume $S(\rho\subtiny{0}{0}{\M{M}_{2}})\geq S(\rho\subtiny{0}{0}{\M{M}_{1}})>0$. Ideal broadcasting then transforms the joint initial state $\rho\subtiny{0}{0}{\M{SM}}=\rho\Sys\otimes\rho\subtiny{0}{0}{\M{M}_{1}}\otimes\rho\subtiny{0}{0}{\M{M}_{2}}$ of system and memory to $\tilde{\rho}\subtiny{0}{0}{\M{S}\M{M}}   =
U\,\rho\Sys\otimes\rho\subtiny{0}{0}{\M{M}_{1}}\otimes\rho\subtiny{0}{0}{\M{M}_{2}}\,U^{\dagger}$\,. The entropy of the initial state is 
\begin{align}
S(\rho\subtiny{0}{0}{\M{SM}})=S(\rho\Sys)
+\sum\limits_{i}S(\rho\subtiny{0}{0}{\M{M}_{i}})
\geq S(\rho\Sys)+2S(\rho\subtiny{0}{0}{\M{M}_{1}}).
\end{align}
For the entropy of the final state, strong subadditivity~\cite{LiebRuskai1973} and subadditivity imply
\begin{align}
    S(\tilde{\rho}\subtiny{0}{0}{\M{SM}})    &\leq\sum\limits_{i}S(\tilde{\rho}\subtiny{0}{0}{\M{SM}_{i}})-S(\tilde{\rho}\Sys)
    \leq\sum\limits_{i}S(\tilde{\rho}\subtiny{0}{0}{\M{M}_{i}})+S(\tilde{\rho}\Sys).
\end{align}
For ideal semiclassical broadcasting, the final-state diagonals of the memories encode the same information as the diagonal of the original system state, and the entropy of the final memory states is upper-bounded by the Shannon entropy of $X$, $S(\tilde{\rho}\subtiny{0}{0}{\M{M}_{i}})\leq H(\{p_{x}\})=H(X)$. 
The final system entropy is bounded by $S(\tilde{\rho}\Sys)\leq\log(d\Sys)$, while unitarity implies $S(\rho\subtiny{0}{0}{\M{SM}})= S(\tilde{\rho}\subtiny{0}{0}{\M{SM}})$. Combining this with the previous inequalities we have
\begin{align}
 S(\rho\Sys)+2\,S(\rho\subtiny{0}{0}{\M{M}_{1}})\,\leq\,2\,H(X)\,\,+\,\log(d\Sys)\,,
 \label{eq:app_ideal broadcasting 2 memories main}
\end{align}
which must hold also when $H(X)=S(\rho\Sys)=0$, which implies $2\,S(\rho\subtiny{0}{0}{\M{M}_{1}})\,\leq\,\log(d\Sys)$. Since $S(\rho\subtiny{0}{0}{\M{M}_{1}})$ is a non-zero but, in principle, arbitrarily small constant, this inequality limits the choice of the initial memory state but does not in itself yield a contradiction.\\[-2.5mm] 

However, since the ideal semiclassical broadcasting has now resulted in two final states $\tilde{\rho}\subtiny{0}{0}{\M{M}_{i}}$ with the same diagonal as the initial system state $\rho\Sys$, we can repeat the ideal semiclassical broadcasting $k$ times, transferring the information from two to $2^{k}$ memories. Applying the same arguments as before, the analogue expression to inequality (\ref{eq:app_ideal broadcasting 2 memories main}) becomes
\begin{align}
 S(\rho\Sys)+2^{k}\,S(\rho\subtiny{0}{0}{\M{M}_{1}})\,\leq\,2^{k}\,H(X)\,\,+\,\log(d\Sys)\,.
 \label{eq:app_ideal broadcasting 2 to the k memories}
\end{align}
The conditions on the semiclassical broadcasting procedure must hold for all $\rho\Sys$, including those states for which $H(X)<S(\rho\subtiny{0}{0}{\M{M}_{1}})$. But for any initial full-rank memory state with fixed  $S(\rho\subtiny{0}{0}{\M{M}_{1}})$ there exists a finite $k\in\mathds{N}$ such that $2^{k}\bigl[S(\rho\subtiny{0}{0}{\M{M}_{1}})-H(X) \bigr]>\log(d\Sys)-S(\rho\Sys)\leq \log(d\Sys)$, violating the inequality, thus showing that ideal semiclassical broadcasting is impossible.
\end{proof}

Let us make two remarks: First, the contradiction already arises for $S(\rho\Sys)=H(X)=0$. Indeed, for pure states an alternative proof is as follows: If $\operatorname{rank}(\rho\Sys)=1$ and two initial memories have $\operatorname{rank}(\rho\subtiny{0}{0}{\M{M}_{1}})=r_{1}\geq d\Sys$ and $\operatorname{rank}(\rho\subtiny{0}{0}{\M{M}_{2}})=r_{2}\geq d\Sys$, respectively, then $\operatorname{rank}(\rho\subtiny{0}{0}{\M{S}\M{M}_{1}\M{M}_{2}})=r_1 r_2 \geq d\Sys^{\,2}$, and unitarity means $\operatorname{rank}(\rho\subtiny{0}{0}{\M{S}\M{M}_{1}\M{M}_{2}})=\tilde{\rho}\subtiny{0}{0}{\M{S}\M{M}_{1}\M{M}_{2}}$. 
Meanwhile, the condition for ideal semiclassical broadcasting (Definition~\ref{defi:ideal broadcasting}) implies that the final states of both memories are also pure, $\operatorname{rank}(\tilde{\rho}\subtiny{0}{0}{\M{M}_{i}})=1$, and hence $\operatorname{rank}(\tilde{\rho}\subtiny{0}{0}{\M{S}\M{M}_{1}\M{M}_{2}})=\operatorname{rank}(\tilde{\rho}\subtiny{0}{0}{S})\leq d\Sys$.\\[-2mm]

Second, the proof also goes through when assuming that the initial ideal semiclassical broadcasting is performed with any $N$, leading to a factor $N^k$ in the final formulas. For $N=1$, an unbiased measurement realizes ideal semiclassical broadcasting.\\[-2mm]

Nevertheless, thermodynamic resources must be invested even in this ideal case. To study the limitations of non-ideal semiclassical broadcasting procedures, it is crucial to characterize these resources. In the next section, we therefore derive a bound{\textemdash}the Holevo{\textendash}Landauer bound{\textemdash}that constrains the amount of information copied to the memory via thermodynamic quantities.


\section{The Holevo{\textendash}Landauer bound}\label{sec:holevo landauer}

{\noindent}To quantify the amount of information copied to the memory we use the Holevo information $\chi$. For an ensemble $\xi\Memo = \{ p_x, \rho\suptiny{0}{0}{(x)}\Memo\}$ of states $\rho\suptiny{0}{0}{(x)}\Memo$ occurring with probabilities $p_x$, it is given by~\cite{Holevo1973}
\begin{align}\label{eq: holevo}
\chi\bigl(\xi\Memo\bigr)
&=\, S(\rho\Memo) - \sum_x p_x S(\rho\suptiny{0}{0}{(x)}\Memo),
\end{align}
and quantifies the maximal amount of information that can be encoded in the state $\rho\Memo=\sum_{x}p_{x}\rho\suptiny{0}{0}{(x)}\Memo$. A quantity that expresses the thermodynamic cost of a physical process is the entropy production $\langle \Sigma \rangle = \beta\Delta Q + \Delta S$, where $\Delta S = S(\trho\Sys)-S(\rho\Sys)$ is the entropy change of the system, and $\Delta Q = \tr(H(\trho\Memo -\rho\Memo))$ can be interpreted as the heat dissipated by the memory. For instance, Landauer's bound~\cite{landauer1961} for bit erasure states that $\langle \Sigma \rangle\geq0$, and the entropy production vanishes only for processes using infinite resources~\cite{TarantoBakhshinezhadEtAl2023}. For general thermodynamic processes, $\langle \Sigma \rangle$ depends on the correlations created between $\M{S}$ and $\M{M}$~\cite{reebandwolf2014}, which can be related to the information encoded in the memory, quantified by $\chi\bigl(\xi\Memo\bigr)$.\\[-1.5mm] 

\begin{Theorems}{Holevo{\textendash}Landauer bound}{HL bound}
For a non-ideal semiclassical broadcasting process $\rho\Sys\otimes\rho\Memo\mapsto\tilde{\rho}\subtiny{0}{0}{\M{SM}}=U\,\rho\Sys\otimes\rho\Memo\,U^{\dagger}$ represented by a joint unitary on the system $\M{S}$ with initial state $\rho\Sys$ and a memory $\M{M}$ initially in a thermal state $\rho\Memo=\tau\Memo=e^{-\beta H\Memo}/\tr(e^{-\beta H\Memo})$ at inverse temperature $\beta$, the entropy production $\langle \Sigma \rangle$ is bounded from below by
\begin{align}
\langle \Sigma \rangle \,\geq\, \chi\bigl(\xi\Memo\bigr) + \beta \Delta F\Memo,
\label{prop. holevo-landauer bound}
\end{align}
where $\chi\bigl(\xi\Memo\bigr)$ is the Holevo quantity with respect to $\xi\Memo = \{ p_x, \rho\suptiny{0}{0}{(x)}\Memo\}$, bounding the accessible information (about the diagonal of $\rho\Sys$ with respect to the chosen basis) in the memory, and $\Delta F\Memo = \Delta E\Memo + \beta^{-1}\Delta S\Memo$ is the free-energy variation of the memory, with $\Delta E\Memo = \tr(H\Memo(\trho\Memo-\rho\Memo))$ and $\Delta S\Memo = S(\trho\Memo)-S(\rho\Memo)$ for $\tilde{\rho}\Memo=\tr\Sys(\tilde{\rho}\subtiny{0}{0}{\M{SM}})$.
\end{Theorems}

\begin{proof}
For the situation we consider, the Reeb-Wolf equality form of Landauer's bound (Theorem 3 in~\cite{reebandwolf2014}) applies,
\begin{align}
         \langle \Sigma \rangle = I(\trho\SM) + D(\trho\Memo||\rho\Memo),
         \label{eq:reeb and wolf}
\end{align}
where the relative-entropy $D(\trho\Memo|\nl|\nr\rho\Memo)=\tr\bigl(\trho\Memo[\log(\trho\Memo)-\log(\rho\Memo)]\bigr)$ can be written as $D(\trho\Memo|\nl|\nr\rho\Memo)=\beta \Delta F\Memo$ for initially thermal states $\rho\Memo=\tau\Memo$, and $I(\trho\SM)=S(\trho\Sys)+S(\trho\Memo)-S(\trho\SM)$ is the mutual information. 
The latter is non-increasing under local CPTP maps, which can be seen by noting that the mutual information is the relative entropy to a product of the reduced states, $I(\trho\SM)=D(\trho\SM|\nl|\trho\Sys\otimes\trho\Memo)$, see, e.g.,~\cite[p.~668]{BertlmannFriis2023}, and the relative entropy is non-increasing under CPTP maps $\Lambda$, $D(\rho\nr|\nl|\nr\sigma) \geq D(\Lambda[\rho]\nr|\nl|\Lambda[\sigma])$, see, e.g.,~\cite[p.~673]{BertlmannFriis2023}, thus $I(\trho\SM)\geq I(\Lambda\Sys\otimes\Lambda\Memo[\trho\SM])$. Applying a local dephasing map with respect to $\{\ket{x}\}_x$ on $\M{S}$ one obtains a state of the form $\sum_{x}p_{x}\ket{x}\!\!\bra{x}\otimes\rho\suptiny{0}{0}{(x)}\Memo$, for which the mutual information matches the Holevo information $\chi\bigl(\xi\Memo\bigr)$.
\end{proof}

Not all information represented by $\chi\bigl(\xi\Memo\bigr)$ can necessarily be extracted through a given measurement due to the potential indistinguishability of the states $\rho\suptiny{0}{0}{(x)}\Memo$. The accessible information $I_{\text{acc}}$ is the maximal information that can be obtained from the ensemble, 
\begin{align}
    I_{\text{acc}}(\xi\Memo)= \max\subtiny{0}{0}{\M{P}\in \text{POVM}} \Big(S(\sum_x p_x \rho^{\M{P}_x}) - \sum_x p_x S(\rho^{\M{P}_x})\Big),
\end{align}
where $\rho^{\M{P}_x} = {M_x \rho M^{\dagger}_x}/{\tr{(\M{P}_x \rho)}}$ is the state conditioned on outcome $x$ for a given POVM with elements ${\M{P}_x}=M_x^{\dagger}M_x$ and measurement operators $M_x$. The Holevo quantity matches the accessible information (thus saturating the Holevo bound) if the states $\rho\suptiny{0}{0}{(x)}\Memo$ are pairwise distinguishable, $\rho\suptiny{0}{0}{(x)}\Memo\rho\suptiny{0}{0}{(y)}\Memo = \delta_{x,y} (\rho\suptiny{0}{0}{(x)}\Memo)^2$, in which case there exists a POVM with elements $\tilde{\M{P}}_x$ such that $\rho^{\tilde{\M{P}}_x} = \rho\suptiny{0}{0}{(x)}\Memo$. 
Similarly, the Holevo{\textendash}Landauer bound is tight if the final joint state has \emph{spectrum broadcast structure} (SBS), 
\begin{align}
    \trho\SM &=\, \sum_x \,p_x\, \ketbra{x}{x}\otimes\rho\suptiny{0}{0}{(x)}\Memo\,,
    \label{eq:SBS}
\end{align}
with orthogonal $\rho\suptiny{0}{0}{(x)}\Memo$~\cite{horodecki2015,le2019PRL}. Then the information is fully copied, $I_{\text{acc}}(\xi\Memo)= \chi\bigl(\xi\Memo\bigr) = H(X)$, which implies $\langle \Sigma \rangle\geq H(X)$. The main difference between the former class of states and SBS is in the existence of off-diagonal elements in the system density matrix, characterized by $\trho_{\mathrm{off}}$: A state resulting from ideal semiclassical broadcasting satisfies SBS if it is diagonal in the chosen measurement basis.


\section{Variants of non-ideal semiclassical broadcasting}\label{sec:variants of semiclassical broadcasting}

\begin{figure*}[t]
  \centering
  \includegraphics[width=0.90\textwidth]{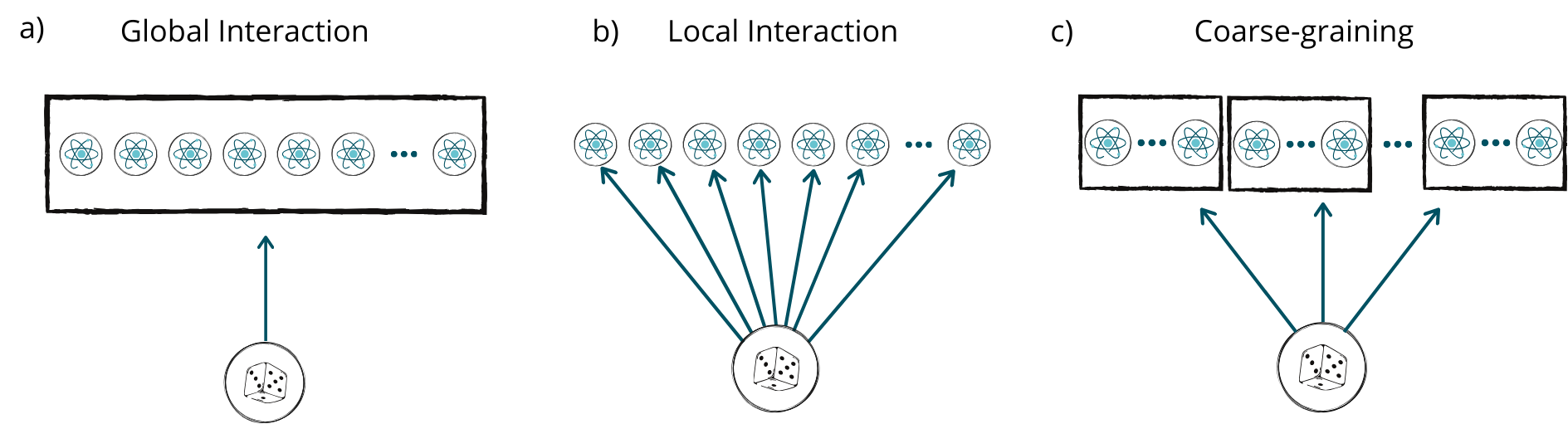}
  \caption{{\bf Different forms of broadcasting to a multi-component memory}. Information can be broadcast to the memory \textbf{a)}~globally, via an interaction Hamiltonian that acts non-trivially on the system and on all memory components, or \textbf{b)}~locally, via consecutive pair-wise coupling of the system to the individual memory components. In the latter case, one of the memory components can contain the full system information or the information can be stored imperfectly in several components individually. However, coarse-graining over different components \textbf{c)}~can lead to recover some if not all of the information.}
  \label{fig: interactions}
\end{figure*}

{\noindent}We have so far seen that ideal semiclassical broadcasting (Definition~\ref{defi:ideal broadcasting}) is not a practical option, as this would require infinite resources (Theorem~\ref{theorem:no-go}) and that the (finite) cost in terms of thermodynamic resources for any practical (and hence non-ideal) semiclassical broadcasting process (Theorem~\ref{theorem:HL bound}) bounds the amount of information copied to the memory. Now, we want to understand in more detail how such non-ideal semiclassical broadcasting could be realized, and in what aspects different non-ideal realizations differ. To gain these insights, we once again start with ideal semiclassical broadcasting, which implies the chain of equalities
\begin{align}
 I_{\text{acc}}(\xi\subtiny{0}{0}{\M{M}_{i}}) =  \chi(\xi\subtiny{0}{0}{\M{M}_{i}}) = H(X) = S(\rho\Sys\suptiny{1}{0}{\mathrm{diag}})\ \forall\,i\,,
 \label{eq: ideal broadcasting}
\end{align}
where $\rho\Sys\suptiny{1}{0}{\mathrm{diag}}$ is the initial state of $\M{S}$ after complete dephasing. 
For a single memory ($N=1$), an unbiased interaction between the system and the single memory satisfies these conditions.

If there are multiple memory components, one could consider a globally unbiased interaction of the system with all memory components, as illustrated in Fig.~\ref{fig: interactions}~a). However, such an interaction will not result in unbiased information in the reduced states of individual components, as demonstrated by Theorem~\ref{theorem:no-go}.\\[-2.5mm]

Alternatively, we could consider successive, individually unbiased local interactions of the system with each of the memory components, as illustrated in Fig.~\ref{fig: interactions}~b). However, as unbiased measurements are invasive for full-rank initial states of the memory, each of these interactions disturbs the system. The diagonal of the final system state contains less information than that of the initial state, $S(\tilde{\rho}\Sys\suptiny{1}{0}{\mathrm{diag}})\geq S(\rho\Sys\suptiny{1}{0}{\mathrm{diag}})$. 
While the very first interactions would hence transfer the information about the diagonal of $\rho\Sys$ to one of the memory components in an unbiased way, the second and every following such interaction would result in unbiased information on ever more disturbed states of the system, containing less and less information about the original system state.\\[-2.5mm] 

For local, non-invasive interactions, the system can be repeatedly probed without disturbance, $S(\tilde{\rho}\Sys\suptiny{1}{0}{\mathrm{diag}})= S(\rho\Sys\suptiny{1}{0}{\mathrm{diag}})=H(X)$, every memory component contains the same information in the end, but the information stored in each component{\textemdash}the probability distribution $\{q_{x}\suptiny{0}{0}{(i)}\}_x$ on the right-hand side of Eq.~(\ref{eq:ideal broadcasting}){\textemdash}does not match the original information $\{p_x\}_x$. Consequently, the operators $\rho\suptiny{0}{0}{(x)}\subtiny{0}{0}{\M{M}_{i}}$ in the decomposition $\rho\subtiny{0}{0}{\M{M}_{i}}=\sum_{x}p_{x}\rho\suptiny{0}{0}{(x)}\subtiny{0}{0}{\M{M}_{i}}$ are not mutually orthogonal and the Holevo bound~\cite{Holevo1973} is not tight, $I_{\text{acc}}(\xi\subtiny{0}{0}{\M{M}_{i}}) \leq  \chi(\xi\subtiny{0}{0}{\M{M}_{i}}) \leq H(X)$.\\[-2.5mm]

As showcased by this non-exhaustive list of examples, different interactions realizing non-ideal semiclassical broadcasting to more than one memory component thus have different advantages and disadvantages, as summarized in Table~\ref{table:info_relations}.\\[-2mm]

\renewcommand{\thetable}{\arabic{table}}
\setlength\extrarowheight{3.5pt}
\begin{smallboxtable}{Comparison of information relations}{table:info_relations}
\begin{center}
\hspace*{-5.0mm}
\begin{tabular}{|c|c|c|}
\hline
     \makecell{SCB variant} & $N$   
     &   information relations    \\ \hline
     \makecell{ideal}       & $\geq 1 $  & $I_{\text{acc}}(\xi\Memo) =  \chi(\xi\Memo) = H(X) \leq S(\trho\Sys\suptiny{0}{0}{\mathrm{diag}}) $ \\ \hline
     \makecell{global\\ unbiased}      & $=1 $  & $I\subtiny{0}{0}{\mathrm{acc}}(\xi\Memo) = \chi(\xi\Memo) = H(X)\leq S(\trho\Sys\suptiny{0}{0}{\mathrm{diag}})$  \\ \hline
     \makecell{local\\ non-invasive}  & $\geq 1 $ &   $I_{\text{acc}}(\xi\Memo)\leq \chi(\xi\Memo) \leq  H(X) = S(\trho\Sys\suptiny{0}{0}{\mathrm{diag}})$   \\ \hline
     \makecell{objectivity}  & $\geq 1$    &   $I_{\text{acc}}(\xi\Memo)= \chi(\xi\Memo) = H(X) = S(\trho\Sys\suptiny{0}{0}{\mathrm{diag}})$    \\ \hline
     \makecell{SBS}  &$ \geq 1$  &   $I_{\text{acc}}(\xi\Memo)= \chi(\xi\Memo) =  H(X) = S(\trho\Sys) $  \\ \hline
\end{tabular}
\end{center}
\small{
The table shows the relation between the accessible information $I_{\text{acc}}(\xi\Memo)$, the Holevo information $\chi(\xi\Memo)$ of the memory, the Shannon entropy $H(X)$ of the stored information, and the von Neumann entropy of the final system state $\trho\Sys$ or its completely dephased version $\trho\Sys\suptiny{0}{0}{\mathrm{diag}}=\sum_x\bra{x}\trho\Sys\ket{x}\ket{x}\!\!\bra{x}$, for ideal (Definition~\ref{defi:ideal broadcasting}), and non-ideal but either unbiased or non-invasive broadcasting. To satisfy the objectivity criterium from~\cite{zurek04}, the condition $I_{\text{acc}}(\xi\Memo)= H(X)$  is required in addition, while spectrum broadcast structure (SBS)~\cite{horodecki2015} implies $\rho\Sys=\trho\Sys$.
}
\end{smallboxtable}
 

In principle it is possible to interpolate between the global and local interactions above by \emph{coarse graining} the memory, that is, by subdividing the memory into $N$ components each consisting of $n$ subsystems. In this scenario, the interactions occurs ``locally" in the sense that the system subsequently interacts with each of the $N$ ($n$-partite) memory components, but each of these interactions ``globally" couples the system to all $n$ subcomponents, as illustrated in Fig.~\ref{fig: interactions}~c). 
Then, even for thermal initial states of the memory, local interactions between the system and and $n$-partite memory component that are either unbiased (and invasive) or non-invasive (and biased) but maximally correlating [see the discussion below Eq.~(\ref{eq:cmax})] have the property that increasing $n$ enhances the faithfulness, with $\cmax\rightarrow1$ and $n\rightarrow\infty$, see, e.g., Fig.~\ref{fig:cmax}. In other words, at fixed initial temperature, more correlations can be created between the post-interaction system state and the memory for larger memories: The individual interactions become more ideal; non-invasive ones become less biased; non-ideal unbiased ones become less invasive, resulting in a more accurate transfer of information. This ensures that the same information can be redundantly transferred to any number $N$ of \emph{macroscopic} memories with $n$ components each, where it can be accessed by different observers.\\[-3mm]

In contrast, for microscopic memories, observers need to employ \emph{local non-invasive interactions} to ensure that all observers receive the same information while preserving the original information in the system. The transferred information is then necessarily biased, but this bias can in principle be removed by post-processing information from different memory components, as we show in the next section. 


\subsection{Reconstruction via post-processing}
\vspace*{-1mm}

{\noindent}Despite the general non-ideality of the information transfer in semiclassical broadcasting, post-processing permits full information recovery. We propose a protocol for the full recovery of the original information using local interactions with $N$ memory components each with $n=d\Sys-1$ subcomponents, given knowledge of the initial temperature and Hamiltonian of the memory.\\[-3mm]

Let us consider a memory with $N$ initially identical components $\M{M}_j$, each comprising (at least, but for simplicity, we assume exactly) $n=d\Sys-1$ subsystems, corresponding to a coarse graining of an $(N\cdot n)$-partite memory, as illustrated in Fig~\ref{fig: interactions}~c). The system $\M{S}$ sequentially interacts with each of the $N$ components via unitaries $U\suptiny{0}{0}{(j)}$ that act locally, i.e., as a joint unitary on $\M{S}\M{M}_j$, but trivially on all $\M{M}_k$ for $k\neq j$. For instance, we can choose a local unitary  $U=\bigotimes_{j=1}^{N} U\suptiny{0}{0}{(j)}$ on the $N$-partite memory, where the $U\suptiny{0}{0}{(j)}$ are controlled unitaries jointly acting on $\M{S}$ and the $j$-th $n$-partite memory component $\M{M}_j$ as 
\begin{align}
    U\suptiny{0}{0}{(j)} &=\,\sum_x \ketbra{x}{x}\Sys\bigotimes_{i=1}^{n} V\suptiny{0}{0}{(i)}_x.
\end{align}
This interaction is non-invasive (but generally not unbiased), meaning the diagonal of the system density operator is not disturbed and each of the sequential interactions can thus access the same system information, but the latter is only imperfectly replicated in each memory component, where the information is represented by the probability vector $q_y = \sum_x p_x\, a_{x,y}$ with $a_{x,y}$ as in Eq.~\eqref{eq: max correlation}.\\[-3mm]

\begin{figure}[t]
  \centering
  \includegraphics[width=0.35\textwidth]{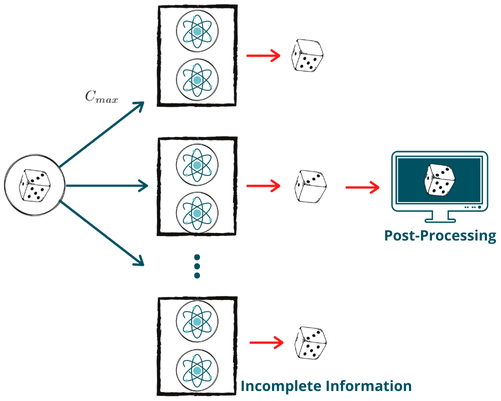}
  \caption{
  {\bf Illustration of reconstruction scheme.} By using detailed information about the initial memory state and Hamiltonian, and using a specific non-invasive coarse-grained interaction that couples the system to $n=d\Sys-1$ subcomponents of any one of the $N$ memory components, one may recover the full information on the diagonal of the original system state by post-processing the information transferred to the $n=d\Sys-1$ subcomponents. 
  }
  \label{fig: post-processing}
\end{figure}

We now wish to recover the full information by post-processing information from the $n=d\Sys-1$ subcomponents of one of the $N$ memories. However, the information copied to each subcomponent is biased. To remove this bias, we need to copy different facets of the information to each subcomponents, as illustrated in Fig.~\ref{fig: post-processing}. To this end, one may notice that there is some freedom in ordering what we refer to here as the anti-correlation terms $a_{x,y}$ for $y\neq x$ satisfying $\sum_{y\neq x}a_{x,y} = 1 - \cmax$, see Eq.~(\ref{eq: max correlation - anti}). It is possible to exploit this freedom of permuting the anti-correlation terms without affecting~$\cmax$. To do so, we choose $V\suptiny{0}{0}{(i)}_x$ to perform a cyclic permutation over the off-diagonal elements of the matrix $A = (a_{x,y})$, resulting in 
$a_{x,y}\suptiny{0}{0}{(i)} = a_{x,(y-i)\operatorname{mod}(d\Sys)},$ for $i\in \M{S}\suptiny{0}{0}{d\Sys}_{x,y}=\{0,\ldots, d\Sys-1 | i \neq (y-x)\operatorname{mod}(d\Sys)\}$, and $\M{S}\suptiny{0}{0}{d\Sys}_{x,y}$ is a set, with cardinality $d\Sys-1$. The sum over all $d\Sys-1$ possible permutations contains all elements of the $x$-th row of the matrix $A$ except for the diagonal element $a_{x,x}$, such that
\begin{align}
    \sum_{i\in\M{S}\suptiny{0}{0}{d\Sys}_{x,y}} a_{x,y}\suptiny{0}{0}{(i)} &=\, \sum_{i\in\M{S}\suptiny{0}{0}{d\Sys}_{x,y}}a_{x,(y-i)} \,=\, \sum_{y\neq x}a_{x,y} \,=\, 1 - \cmax\,. 
\end{align}
The information copied to the $i$-th memory subcomponent is then represented by a probability distribution with elements
\begin{align}
        q_y\suptiny{0}{0}{(i)} &=\,  \sum_x p_x\, a_{x,y}\suptiny{0}{0}{(i)} \,=\, 
        \sum_x p_x a_{x,(y-i)} \ \  \forall\, i\in\M{S}\suptiny{0}{0}{d\Sys}_{x,y}\,.
\end{align}
We can then take the average of the probabilities of the different memory subcomponents: an average over all permutations, which yields
\begin{align}
        q_y\suptiny{0}{0}{\text{av}} &=\, \frac{1}{d\Sys-1}\sum_{i} q_y\suptiny{0}{0}{(i)} =  \frac{1}{d\Sys-1}\sum_{i} \sum_x p_x\, a_{x,y}\suptiny{0}{0}{(i)} \nonumber\\
        &=\, \frac{1}{d\Sys-1}\sum_{i} p_y a_{y,y} + \sum_{x\neq y} p_x \sum_{i\in\M{S}\suptiny{0}{0}{d\Sys}_{x,y}}a_{x,y}\suptiny{0}{0}{(i)}\nonumber\\
        &=\, p_y \,\cmax + \frac{1}{d\Sys-1}\sum_{x\neq y} p_x \,(1-\cmax), \nonumber\\  
     &=\, p_y \,\cmax + \frac{1-\cmax}{d\Sys-1}(1-p_y),
\end{align}
since $a_{y,y}=\cmax$, and $\sum_{x\neq y} p_x  = 1-p_y$. 

Therefore, if the memory Hamiltonian and temperature are known, it is possible to obtain $\cmax$, allowing us to fully characterize $p_y$ by post-processing the information copied to $d\Sys-1$ memory components. Detailed knowledge of and control over the memory, combined with the average information copied to different memory subcomponents thus allows one to determine the information to be copied to the memory, even if each component contains biased information. Therefore, with sufficiently many memory components, the global bias can be entirely determined through post-processing. 


\section{The emergence of objectivity}\label{sec:classical limit and emergence of objectivity}

{\noindent}As briefly mentioned already at the end of Sec.~\ref{sec:variants of semiclassical broadcasting}, a different kind of information recovery is possible when the Hilbert-space dimension of each of the $N$ memory components diverges, for instance, when each component consists of infinitely many identical systems at the same fixed temperature. In this limit, $\cmax \rightarrow 1$ in Eq.~\eqref{eq:cmax}, implying that interactions can become unbiased and non-invasive asymptotically despite the initial memory being in a state of full rank~\cite{GuryanovaFriisHuber2018}, thus allowing successive exact copying of the information without disturbing it{\textemdash}\emph{ideal broadcasting}. In this section, we examine this situation more closely using the example from Sec.~\ref{sec:non-ideal measurements}, which leads us to a discussion of the relation of ideal broadcasting to models relevant to studying the emergence of objectivity.\\[-2mm] 


If all correlation terms in Eq.~\eqref{eq: max correlation} approach one ($a_{x,x} = \cmax \rightarrow 1$) for all $N$ memory components coarse-grained over the respective $n$ subsystems, then the anti-correlation terms must tend to zero, $\sum_{y=1}^{d\Sys-1} a_{x,x \oplus y} = 0$. This condition implies that the initial state of each memory component has support in the symmetric subspace, which implies that each memory is rank-deficient, i.e., $\text{rank}(\rho\Memo\suptiny{0}{0}{(j)}) = d\Memo\suptiny{0}{0}{(j)} / d\Sys < d\Memo\suptiny{0}{0}{(j)}$. Therefore, the contribution in each subspace of the initial memory state will be
\begin{align}
    A_{x,x \oplus y}^{(j)} &=\,  V_x \,A_{0,y}^{(j)}\, V_x^{-1}\, \delta_{y,0}\,.
    \label{eq: app_M_N_infity}
\end{align}
As the matrices $A_{x,x}$ have orthogonal support, spanned by the projectors $\Pi_x$, with $\tr(A_{x,x} \Pi_y) = \cmax \delta_{x,y}$, the state of the $j$-th memory component is described by a convex combination of orthogonal states, i.e.,  $\rho\Memo\suptiny{0}{0}{(j)} \,=\, \sum_x \,p_x\, A_{x,x}\suptiny{0}{0}{(j)}$. 
The global state of the system and the $N$ memories is then in the ideal broadcasting form, 
\begin{align}
    \trho_{\mathcal{S} \mathcal{M}_1 \cdots \mathcal{M}_N} &=\, \sum_{x} p_x\, \ketbra{x}{x}\, \bigotimes_{j=1}^N \,A_{x,x}\suptiny{0}{0}{(j)} \,+\, \trho_{\mathrm{off}}.
    \label{eq: ideal broadcasting state}
\end{align}
One can verify that such states satisfy Definition~\ref{defi:ideal broadcasting}.\\[-2mm]

But does such a state lead to the emergence of objectivity? To answer this, we note that the key distinction between this class of states and SBS lies in the presence of off-diagonal elements in the system density matrix, represented by $\trho_{\mathrm{off}}$. In other words, an ideal-broadcasting state corresponds to SBS if it is diagonal in the chosen measurement basis. However, the state in Eq.~\eqref{eq: ideal broadcasting state} is not an SBS state in general due to the presence of additional off-diagonal elements. Nevertheless, objectivity emerges as the system and the memory components share the same information originally encoded in the diagonal of $\rho\Sys$~\cite{horodecki2015}. As expressed via the mutual information, we have
\begin{align}
    I(\trho\SMi) &=\, S(\trho\Sys)+S(\trho\Mindex{i})-S(\trho\SMi)\nonumber\\[1.5mm]
    &=\, S(\trho\Sys)+\bigl[H(X)+\sum_x \,p_x\, S(V_x \rho\Mindex{i} V_x^{-1} )\bigr]\nonumber\\
    &\ \ -\,\bigl[S(\rho\Sys)+S(\rho\Mindex{i})\bigr]
    \,=\, H(X), 
\end{align}
where we have used Eq.~\eqref{eq: app_M_N_infity} in the first step, along with the orthogonality of the ensemble of states $S(\sum_x \,p_x\, V_x\, \rho\Memo\, V_x^{-1} ) = H(X)+\sum_x \,p_x\, S(V_x\, \rho\Memo\, V_x^{-1} )$, and the invariance of the von Neumann entropy under unitaries. In the second step, we notice that $\sum_x\, p_x\, S(V_x\, \rho\Memo\, V_x^{-1} )) = S(\rho\Memo)$ and $S(\rho\Sys) = S(\trho\Sys)$ as the state of the system remains undisturbed.\\[-2mm]

Ideal broadcasting implies information encoding with sufficient redundancy: Observables on different memory components can access the information independently. Different observers arrive at the same conclusion, allowing for the emergence of objectivity~\cite{horodecki2015}. Yet, objectivity criteria as those in~\cite{zurek04, horodecki2015} represent stricter constraints than ideal broadcasting as the latter permits disturbance of the original system state in the form of decoherence of the off-diagonal elements (see Table~\ref{table:info_relations}). Thus, ideal information broadcasting generalizes the concept of objectivity. Realistic non-ideal measurements performed with finite resources hence do not match these objectivity criteria either, even if, as we have shown, coarse-graining and certain large-memory limits allow satisfying them in principle. On the one hand, this ties in with similar findings regarding the interplay of thermality and objectivity in other work~\cite{le2021}. On the other hand, it may be worth considering the merit of a non-ideal notion of objectivity, where agreement among observers regarding a measurement outcome may outweigh concerns for the outcome statistics relative to the state of the measured system.


\section{Discussion}\label{sec:discussion}

{\noindent}We have analysed the problem of semiclassical broadcasting, that is, transferring classical information encoded in the diagonal of a quantum state to several memory components using finite thermodynamic resources. Specifically, we investigate whether it is possible to redundantly copy measurement statistics (classical information) from a quantum system to multiple memory subsystems, given that the memories are initially in mixed states (e.g., thermal states). This problem, which is related to but distinct from both the original broadcasting problem~\cite{BarnumCavesFuchsJozsaSchumacher1996} and the problem of non-ideal projective measurements~\cite{GuryanovaFriisHuber2018}, is motivated by the need to establish objective measurement outcomes in quantum mechanics, where multiple observers must agree on the measurement results without disturbing the original system. The key issue is that finite resources (e.g., finite energy, finite temperature, or finite memory size) prevent the preparation of memories in pure states, which are necessary for ideal broadcasting. This limitation raises the question, whether redundant encoding of classical information (i.e., semiclassical broadcasting) is possible under such constraints, and if so, what are the fundamental limits imposed by thermodynamics?\\ 

We have shown that creating redundant information (i.e., exact copying of classical information to many memory components) is impossible in the situation of full-rank (e.g., thermal) initial memory states and finite dimensions. We have introduced a thermodynamic bound that we call the Holevo{\textendash}Landauer bound, which quantifies the trade-off between information redundancy and thermodynamic cost. The bound shows that increased redundancy (i.e., semiclassical broadcasting to more memory components) is more expensive thermodynamically because more entropy is produced. 

Despite these limitations, we were successful in demonstrating that the original probability distribution can be reconstructed by post-processing the information of multiple memory components. Furthermore, we argue that in the limit where each memory component has infinite dimension, the semiclassical broadcasting process becomes ideal, allowing the original information to be copied perfectly.

Our results contribute to the discussion of the emergence of objectivity in quantum measurements in the classical limit, despite the imperfections introduced by the laws of thermodynamics, as previously discussed in Ref.~\cite{le2021}.

Our work further sheds light on quantifying memory capacity as a thermodynamic resource, connecting to recent results~\cite{milegu2019a}, which demonstrate that the asymptotic thermodynamic capacity of the memory (as captured by the channel capacity of the thermal operation that records the information in the memory) is equal to the free energy of the memory. 


\begin{acknowledgments}
We are grateful to Pharnam Bakhshinezad, Felix C. Binder, Mile Gu, Maximilian P. E. Lock, Fernando de Melo, Emanuel Schwarzhans, and Jake Xuereb for fruitful discussions, and we thank Yog-Sothoth for insights into space, time, and gates. 
T.D. acknowledges support from the {\"O}AW-JESH-Programme and the Brazilian agencies CNPq (Grant No. 441774/2023-7, 200013/2024-6 and 445150/2024-6) and INCT-IQ through the project (465469/2014-0). 
N.F. acknowledges support from the Austrian Science Fund (FWF) through the projects P~31339-N27 and P 36478-N funded by the European Union - NextGenerationEU, as well as from the Austrian Federal Ministry of Education, Science and Research via the Austrian Research Promotion Agency (FFG) through the flagship project HPQC (FO999897481) and the project FO999921407 (HDcode) funded by the European Union – NextGenerationEU.
M.H acknowledges funding from the FQXi Grant Number: FQXi-IAF19-07 from the Foundational Questions Institute Fund, a donor advised fund of Silicon Valley Community Foundation, and from the European Research Council (Consolidator grant ’Cocoquest’ 101043705).
N.F. and M.H. acknowledge funding by the Austrian Federal Ministry of Education, Science and Research via the Austrian Research Promotion Agency (FFG) through the project FO999914030 (MUSIQ) funded by the European Union{\textemdash}NextGenerationEU. 
This publication was made possible through the support of Grant 62423 (``Emergence of Objective Reality: From Qubit to Oscilloscope") and Grant 62179 (``Bridging physical theories: how large-scale laws emerge from the perspective of algorithmic information theory and quantum many-body physics") from the John Templeton Foundation. The opinions expressed in this publication are those of the author(s) and do not necessarily reflect the views of the John Templeton Foundation. 
No Capybaras were harmed as a result of this research project.
\end{acknowledgments}


%


\end{document}